\documentclass[12pt,preprint]{aastex}

\newcommand{\bdv}[1]{\mbox{\boldmath$#1$}}

\def\au{{\rm AU}}

\def\min{{\rm min}}
\def\rel{{\rm rel}}

\def\eff{{\rm eff}}

\def\e{{\rm E}}
\def\bpi{{\bdv\pi}}
\def\bmu{{\bdv\mu}}

\def\bv{{\bf v}}

\begin{document}
\title{The KMTNet/K2-C9 ({\it Kepler}) Data Release}

\author{\textsc{
H.-W. Kim$^{1}$, 
K.-H. Hwang$^{1}$,  
D.-J. Kim$^{1}$, 
M. D. Albrow$^{2}$,
S.-M. Cha$^{1,3}$, 
S.-J. Chung$^{1,4}$, 
A. Gould$^{1,5,6}$, 
C. Han$^{7}$, 
Y. K. Jung$^{8}$, 
S.-L. Kim$^{1,4}$, 
C.-U. Lee$^{1,4}$,
D.-J. Lee$^{1}$,
Y. Lee$^{1,3}$, 
B.-G. Park$^{1,4}$,
R. W. Pogge$^{5}$ 
Y.-H. Ryu$^{1}$, 
I.-G. Shin$^{8}$, 
Y.~Shvartzvald$^{9,^{\dag}}$, 
J. C. Yee$^{8}$, 
W.~Zang$^{10,11}$, 
W. Zhu$^{12}$, 
\\
(KMTNet Collaboration)\\}}

\affil{$^{1}$Korea Astronomy and Space Science Institute, Daejon
34055, Korea}

\affil{$^{2}$University of Canterbury, Department of Physics and
Astronomy, Private Bag 4800, Christchurch 8020, New Zealand}

\affil{$^{3}$School of Space Research, Kyung Hee University,
Yongin, Kyeonggi 17104, Korea}

\affil{$^{4}$Astronomy and Space Science Major, Korea University of
Science and Technology, Daejeon 34113, Korea}

\affil{$^{5}$Department of Astronomy, Ohio State University, 140 W.
18th Ave., Columbus, OH 43210, USA}

\affil{$^{6}$Max-Planck-Institute for Astronomy, K\"{o}nigstuhl 17,
69117 Heidelberg, Germany}

\affil{$^{7}$Department of Physics, Chungbuk National University,
Cheongju 28644, Republic of Korea}

\affil{$^{8}$Harvard-Smithsonian CfA, 60 Garden St.,Cambridge, MA 02138, USA}

\affil{$^{9}$Jet Propulsion Laboratory, California Institute of
Technology, 4800 Oak Grove Drive, Pasadena, CA 91109, USA}

\affil{$^{10}$Physics Department and Tsinghua Centre for
Astrophysics, Tsinghua University, Beijing 100084, China}

\affil{$^{11}$Department of Physics, Zhejiang University, Hangzhou,
310058, China}

\affil{$^{12}$Canadian Institute for Theoretical Astrophysics, 
University of Toronto, 60 St George Street, Toronto, ON M5S 3H8, Canada}

\affil{$^{\dag}$NASA Postdoctoral Program Fellow}

\vfil\eject

\begin{abstract}
We present Korea Microlensing Telescope Network (KMTNet)
light curves for microlensing-event candidates 
in the {\it Kepler} K2 C9 field having peaks within 3 effective
timescales of the {\it Kepler} observations.  These include
181 ``clear microlensing'' and 84 ``possible microlensing'' events
found by the KMTNet event finder, plus 56 other events found by
OGLE and/or MOA that were not found by KMTNet.  All data for the
first two classes are immediately available for public use without restriction.

\end{abstract}


\section{{Introduction}
\label{sec:intro}}

After {\it Kepler} lost its second reaction wheel, 
\citet{gouldhorne} proposed that
it should become a ``microlens parallax satellite'', a concept first advanced
by \citet{refsdal66} a half-century earlier and further articulated by
\citet{gould94}.  By observing the same event simultaneously from two
platforms separated by of order 1 AU, one can measure the projected
velocity, basically the ratio of the lens-source relative proper motion,
$\bmu_\rel$, 
to their relative parallax, $\pi_\rel$,
\begin{equation}
\tilde\bv \equiv {\au}{\bmu_\rel\over\pi_\rel}.
\label{eqn:projvel}
\end{equation}
By itself, this quantity is a powerful constraint on the otherwise
poorly determined lens properties \citep{han95,21event,zhu17a}.
Moreover, if the Einstein radius $\theta_\e$ is also measured
(as is often the case for binary and planetary events), then the
projected velocity $\tilde\bv$, or equivalently the microlens parallax,
\begin{equation}
\bpi_\e \equiv {\pi_\rel\over \theta_\e}{\bmu_\rel\over\mu_\rel},
= {\au\over t_\e}{\tilde\bv\over \tilde v^2},
\label{eqn:bpie}
\end{equation}
enables measurements of both $\pi_\rel=\theta_\e\pi_\e$ and the
lens mass $M=(4c^2/G)\tilde v t_\e\theta_\e$ \citep{gould92}.
Here $t_\e = \theta_\e/\mu_\rel$ is the Einstein timescale.

Up until the time of the proposal by \citet{gouldhorne}, 
there had been only one successful
microlens parallax measurement, which had been carried out by
{\it Spitzer} toward the Small Magellanic Cloud \citep{smc001}.
Eventually {\it Spitzer} would observe many hundreds of microlensing
events \citep{prop2013,prop2014,prop2015a,prop2015b,prop2016}.
However, the point that we wish to emphasize here is the
fundamental difference between {\it Spitzer} and {\it Kepler} microlensing.
With its narrow-angle camera {\it Spitzer} must point at individual 
microlensing events, and hence an event must be recognized as interesting
before {\it Spitzer} can start observing it.  
Moreover, due to operational constraints,
{\it Spitzer} cannot begin observing an event until 3--9 days after
the event is singled out for observations (Figure~1 of \citealt{ob140124}).
On the other hand,  {\it Spitzer} has the advantage that it can be
pointed at any microlensing event over the $\sim 100\,{\rm deg}^2$
field in which they are discovered.

By contrast, {\it Kepler} must observe continuously for many weeks
without receiving instructions from Earth and so cannot be pointed
at any particular event (with the exception of a few extremely long ones).
On the other hand, it does observe over a wide enough
field that a significant number of stars within this field undergo
microlensing events.  It therefore observes many
microlensing events without specific targeting, and
thus independently of the discovery of these events from the ground.
Unfortunately, although
the full {\it Kepler} field is $\sim 100\,{\rm deg}^2$, memory constraints
restrict the contiguous area of observations
to $\sim 4\,{\rm deg}^2$.  Nevertheless,
this area can be chosen to contain some of the densest microlensing
fields.  See \citet{henderson16}.

These different engineering characteristics lead to very different
regimes of high science performance.  In particular, {\it Kepler}
is far better suited to the study of very short timescale events,
especially those due to free-floating planets (FFPs) because these
are typically over before {\it Spitzer} could target them.  Hence,
study of (or constraints upon) FFPs were major goals for the {\it Kepler}
K2 C9 campaign, which operated from April 22.596 to July 2.936 2016.

It was recognized from the beginning, both by NASA and the microlensing
community, that the analysis of {\it Kepler} microlensing data
would be exceptionally difficult because the microlensing fields
are extremely crowded while {\it Kepler} has $4^{\prime\prime}$ pixels.
Moreover, in its two-wheel mode, {\it Kepler} oscillates on a 
not-quite-repeating 6-hour cycle.  
Nevertheless \citet{zhu17b} solved many of the photometry problems 
and applied these solutions to several different events
\citep{zhu17b,zhu17c,ob161190}.
However, the \citet{zhu17b} approach requires a lot of tender loving
care (TLC), and this increases the premium on identifying events
that are likely to have scientific importance.

{\it Kepler} K2 C9 microlensing was strongly supported by several
microlensing surveys.  
The Optical Gravitational Lensing Experiment \citep{ews1}
(OGLE)\footnote{http://ogle.astrouw.edu.pl/ogle4/ews/ews.html}
altered their observing strategy so that the cadence of 
all five of their fields that strongly overlap
the K2 C9 superstamp were observed at OGLE's maximum rate,
$\Gamma=3\,{\rm hr}^{-1}$, during the K2 C9 observations.  OGLE also
identified 54 long timescale events within the K2 C9 field but
outside the superstamp, for individual K2 C9 observations.
The Microlensing Observations in Astrophysics \citep{ob03235}
(MOA)\footnote{http://www.massey.ac.nz/$\sim$iabond/moa/alert2016/alert.php}
collaboration also altered its observing strategy to support K2 C9.
Six MOA fields overlap the superstamp, four of which are usually
observed at the highest cadence $\Gamma=4\,{\rm hr}^{-1}$
but one at very low cadence. This field was upgraded to
$\Gamma=1.2\,{\rm hr}^{-1}$.

In addition, there were several special surveys devoted to the K2 C9
field,  The first two of these were both in Hawaii.  
Using the Canada-France-Hawaii Telescope (CFHT), the 2016
CFHT-K2C9 Multi-color Microlensing Survey \citep{cfht}
conducted an independent
survey in the $g$-, $r$-, and $i$-bands, which together cover almost
the entire {\it Kepler} bandpass. The CFHT data can be used to predict
the magnitude in the Kepler bandpass $K_p$, which, has an uncertainty
of ∼ 0.02 mag in the $(K_p - r)$ color for typical microlensing events.

The United Kingdom Infrared Telescope (UKIRT) survey \citep{ukirt}
observed the entire K2 C9 superstamp in the near infrared. The 
observations covered 91 nights from April 8 to July 8, with a nominal
cadence of 2--3 per night.  The data are available at
https://exoplanetarchive.ipac.caltech.edu/docs/UKIRTMission.html .

The LCOGT Network \citep{lcogt} conducted a survey using 1m telescopes
equipped with $26^\prime\times 26^\prime$ cameras at three sites
in $i$-band.  From Australia, they observed 19 fields in the K2 C9
superstamp, while from Chile and South Africa they observed 10 fields
containing high magnification events, from both the superstamp and
the additional events mentioned above.

The Korea Microlensing Telescope Network (KMTNet, \citealt{kmtnet})
supported the K2 C9 campaign in three ways.  First, we altered our
observing strategy to temporarily increase the cadence on the K2 C9
superstamp.  This superstamp is covered by two KMTNet fields
BLG02 and BLG03 (together with their slightly offset counterparts
BLG42 and BLG43).  Each KMTNet field is $4\,{\rm deg}^2$.
Under our standard observing protocol for
2016, these two fields were observed at a cadence 
$\Gamma=4\,{\rm hr}^{-1}$ from each of our three observatories, in
Chile (KMTC), South Africa (KMTS), and Australia (KMTA).  However,
for all but the final 15 days of K2 C9 (during which {\it Spitzer}
microlensing began), we increased the cadence at KMTS and KMTA
to $\Gamma=6\,{\rm hr}^{-1}$.  That is, the special K2 observing protocol
was in force from April 23 to June 16.

Second, we altered our data policy for all events that are independently
discovered by KMTNet and that peak in or near the K2 C9 window.
KMTNet generally
aims to make all of our microlensing light curves publicly available
in a timely fashion \citep{eventfinder}.  However, as discussed in
Section~\ref{sec:policy}, we have further relaxed this policy
for KMTNet-discovered K2 C9 events.

Third, for the special case of K2 C9 events, we are also making
publicly available all KMTNet data for microlensing events found by
other groups (in practice, OGLE and MOA), but that were not
independently identified by KMTNet.  In this case, however,
we require that prospective users of these data obtain permission
from the discovering group(s) before publishing them.

\section{{2016 Event Finder}
\label{sec:eventfinder}}

The events presented here were found by applying the KMTNet event finder
to 2016 data that overlap the K2 C9 field and then excluding those
events that peak well outside the K2 C9 window.  The KMTNet event finder
was described in detail by \citet{eventfinder}.  In brief, each light curve
is fit to a dense grid of point-lens models, parametrized by 
$(t_0,t_\eff,u_0)$, where $t_0$ is the peak time, $t_\eff$ is the effective
timescale, and $u_0$ is the impact parameter.  Because the fit is
carried out only on the interval $t_0\pm 5\,t_\eff$, the $u_0$ sampling
can be reduced to just two values (0 and 1), yet still remain ``dense''
\citep{eventfinder}.  Those events that pass a $\chi^2$ threshold are
cataloged.  Then these are grouped by a ``friends-of-friends'' algorithm.
The output is shown to an operator in several different displays, who
then classifies them as ``clear microlensing'', ``possible microlensing'',
or as one of several classes of variables or artifacts.

Here, we describe in greater detail only the changes in the 2016
version of the algorithm relative to 2015.  The most important change
is that data from all three observatories are fitted simultaneously to
the same model.  (In 2015, by contrast, only the KMTC data were fitted.
Then, only if the event passed the $\chi^2$ threshold
and was subsequently selected by the operator, would the KMTS and KMTA
data be reduced and inspected.)

Second, for the cases that the same star is observed in two 
or more KMTNet fields,
we fit all the light curves simultaneously when possible.  This was
not necessary for 2015 because there were no overlapping fields.  In
2016, however, fields (BLG01, BLG02, BLG03) were observed slightly
offset as (BLG41, BLG42, BLG43).  See Figure~12 of \citet{eventfinder}.
In addition, BLG02 and BLG03 have a quite significant overlap (roughly
$0.4\,{\rm deg}^2$), which lies almost entirely in the K2 C9 superstamp.  
Thus, there
can up to four overlapping fields, in which case we would fit 12
light curves simultaneously.  (In fact, there are very occasionally
five overlapping fields).  However, we are able to unambiguously
identify two catalog stars from different fields as being the
``same star'' only for regions covered by the OGLE-III star catalog
\citep{oiiicat}.
That is, for these regions, our input catalog is derived from OGLE-III,
and so is identical for all fields.  However, for regions not covered
by OGLE-III, for which our star catalog is derived from DoPhot \citep{dophot}
analysis of KMTNet images, we must analyze the light curves from different
fields independently (although the light curves from different observatories
are still analyzed jointly).

Third, we set a $\chi^2>500$ threshold for all stars that lie
in one of the six fields, BLG01, BLG02, BLG03, BLG41, BLG42, BLG43,
and that lack OGLE-III counterparts.  For the remaining stars
(roughly 95\% of the total), i.e.,
those lying in these six fields that have OGLE-III counterparts
as well as all stars that lie in other fields, we adopt $\chi^2>1000$.  

The reason for using two different $\chi^2$ thresholds
is that light curves in the first category have potentially 
confirming information from other fields, while those in the second
do not.

Fourth, while \citet{eventfinder} expressed the hope that the
event finder could be applied to very short effective timescales
in 2016, in fact this proved impractical due to an excess of spurious
candidate events.  Hence, we adopt $t_\eff\geq 1\,$day.  This is
a significant shortcoming, particularly for the K2 C9 field, for
which FFPs are especially important.  We are currently working
on an alternate algorithm for very short events.  If successful,
we will also release the data for those events.  However, we do
not wish to delay the present release.

Fifth, we cross-checked all candidates against variables and artifacts
that were identified in 2015 as well as against all published variables
that we could identify.  However, because the field locations (and
so star catalogs) changed between 2015 and 2016, we could cross-check
against 2015 only for OGLE-III catalog stars.

Sixth, for each event we inspected the combined 2016+2017 light curve,
thereby removing 49 variables, mostly cataclysmic variables (CVs),
but also other episodic repeaters as well as a few long-period variables.

Finally, we note that several improvements were made during the 2016 analysis.
These all occurred after the K2 C9 superstamp was processed and so apply
only to other fields, but we briefly mention them here.  A complete
description will be given in the 2016 season data release paper.
First, we created a new periodic-variable finder, which was applied
to the candidates (after being grouped).  Second we created an algorithm
to identify an important class of short-timescale artifacts, which was
also applied to grouped candidates in 2016 but which is being incorporated
directly into the event finder for 2017 and future years.

\section{{K2 C9 Selection}
\label{sec:k2c9}}

Light curves are identified for publication within this K2 C9 release 
provided
that they meet two criteria.  First, they must lie in the K2 C9 field
as determined by the \texttt{K2fov} algorithm, which is available at
https://keplerscience.arc.nasa.gov/software.html .
This includes both events in
the superstamp and any of the 37 (out of 54) ``postage stamps'' that
overlap the KMTNet fields.
Second, $t_0$ must lie in the range,
\begin{equation}
t_{{\rm K2C9},-} - 3\,t_\eff < t_0 < t_{{\rm K2C9},+} + 3\,t_\eff ,
\label{eqn:k2c9}
\end{equation}
where $t_0$ and $t_\eff$ are the best-fit values as determined by the
event finder and $t_{{\rm K2C9},\pm}$ are the start and end times of the 
K2 C9 campaign.

Our supplementary release of KMTNet data for events found by other groups
is defined as follows.  First, it is restricted to events that were not
found by the KMTNet event finder for 2016 as a whole.  Second, the event must
lie in the K2 C9 field as determined by the \texttt{K2fov}
algorithm.  Third, it
must satisfy Equation~(\ref{eqn:k2c9}) but with $t_0$ and $t_\eff$ derived
from the fit announced by OGLE (if OGLE found it) or, if not, by the
discovery group.  The value of $t_0$ is simply the one that is reported,
while $t_\eff\equiv \min(u_0,1)t_\e$, where $u_0$ and $t_\e$ are the
reported impact parameter and Einstein timescale, respectively.

These prescriptions should include the overwhelming majority
of events that have useful {\it Kepler} data.  However, if there
are others (and if reasonable scientific justification is given),
we will provide data for these upon request.

\begin{figure}
\plotone{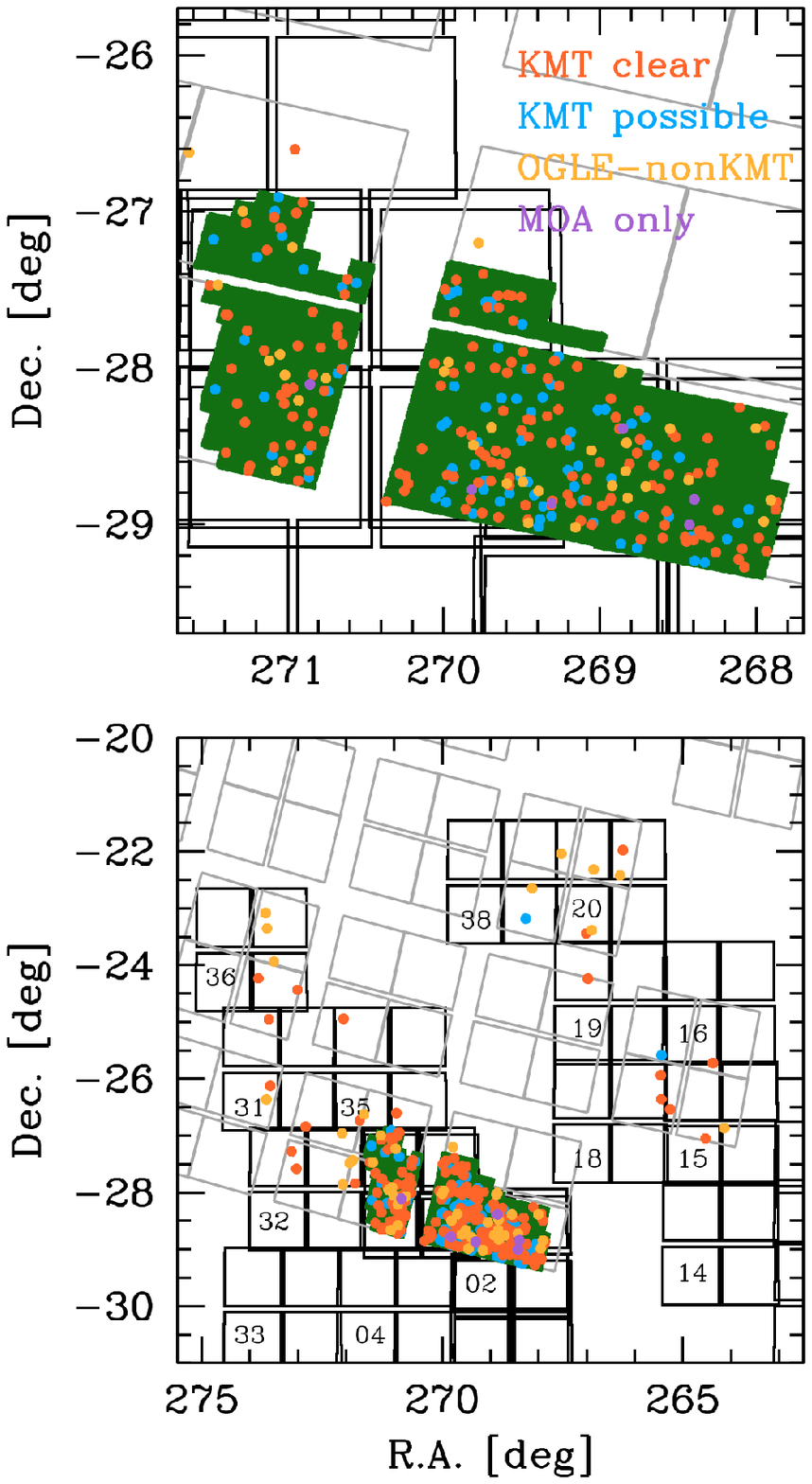}
\caption{Microlensing events with KMTNet data in the {\it Kepler} K2 C9 
field.  The main superstamp is shown in green.  Events that were
independently identified as ``clear'' or ``possible'' microlensing
are shown in red and blue, respectively.  Events found by OGLE but not
KMTNet are shown in gold.  Those found only by MOA are shown in purple.
The active chips of the {\it Kepler} camera are shown in gray outline, while
the KMTNet fields are shown in black outline.  When there is sufficient
space, the KMTNet field number is given in the lower-left chip.  See Figure~12
of \citet{eventfinder} for a more complete representation of these fields.
}
\label{fig:k2kmt}
\end{figure}

Figure~\ref{fig:k2kmt} shows the sky positions of the events with data that
are being released.  The red and blue points indicate ``clear''
and ``possible'' microlensing events found by the KMTNet event finder, 
respectively. The gold points are events found by OGLE but not by KMTNet.
The purple points are events found by MOA only.

\section{{Data Products}
\label{sec:product}}

The data are available at the website 
http://kmtnet.kasi.re.kr/ulens/
with subpages for this paper at
http://kmtnet.kasi.re.kr/ulens/event/2016k2/
and\hfil\break\noindent
http://kmtnet.kasi.re.kr/ulens/event/2016nonkmt/ .  
For each event
that was found independently by KMTNet
(181 ``clear'', 84 ``possible''), there
are two reductions.  The first is the difference imaging analysis (DIA)
used to identify the events, which is derived from publicly available
\citet{wozniak2000} code.  The second is a pySIS reduction employing
the code of \citet{albrow09}.  In most cases, the latter is substantially
better, but it can fail completely in some cases.  
These can easily be identified
by the pictorial representations on the webpage.  We note that 
it may in principle be possible to recover from these failures using
a TLC approach, but we do not generically attempt to do so and such
TLC light curves are not part of this release.

The pages also contain diagnostic information similar to that
of \citet{eventfinder}, i.e.,
finding charts, diagnostic images,
parameters from the fit, as well as the identifier names of
events found by other surveys.

For the 56 events that were not independently found by KMTNet, the pages contain
pySIS reductions but not DIA reductions.

\section{{Data Policy}
\label{sec:policy}}

For K2 C9 events found by KMTNet, there
is no restriction whatever on the use of the released light curves,
other than to cite this paper.  For K2 C9 events found by
other groups, but not found by KMTNet, use of the KMTNet data requires
consent of the discovery group(s).

\acknowledgments 
Work by WZ, YKJ, and AG were supported by AST-1516842 from the US NSF.
WZ, IGS, and AG were supported by JPL grant 1500811.
This research has made use of the KMTNet system operated by the Korea
Astronomy and Space Science Institute (KASI) and the data were obtained at
three host sites of CTIO in Chile, SAAO in South Africa, and SSO in
Australia.
Work by C.H. was supported by the grant (2017R1A4A101517) of
National Research Foundation of Korea.
Work by YS was supported by an appointment to the NASA Postdoctoral
Program at the Jet Propulsion Laboratory, California Institute of
Technology, administered by Universities Space Research Association
through a contract with NASA.

\begin{thebibliography}{99}


\bibitem[Albrow et al.(2009)]{albrow09}Albrow, M.\ D., Horne, K., Bramich, D.\ M., et al.\ 2009, \mnras, 397, 2099














\bibitem[Bond et al.(2004)]{ob03235} Bond, I.A., Udalski, A., Jaroszy\'nski, M. et al.\ 2004, \apj, 606, L155


\bibitem[Brown et al.(2013)]{lcogt}Brown, T.M., Baliber, N., Bianco, F.B. et al. 2013, \pasp, 125, 1031

\bibitem[Calchi Novati et al.(2015)]{21event}  Calchi Novati, S., Gould, A., Udalski, A., et al., 2015, \apj, 804, 20





\bibitem[Dong et al.(2007)]{smc001} Dong, S., Udalski, A., et al. 2007, \apj, 664, 862





\bibitem[Gould(1992)]{gould92} Gould, A. 1992, \apj, 392, 442


\bibitem[Gould(1994)]{gould94} Gould, A. 1994, \apjl, 421, L75




\bibitem[Gould \& Horne(2013)]{gouldhorne} Gould, A. \& Horne, K. 2013, \apjl, 779, L28




\bibitem[Gould et al.(2013)]{prop2013} Gould, A., Carey, S., \& Yee, J. 2013, spitz.prop 10036

\bibitem[Gould et al.(2014)]{prop2014} Gould, A., Carey, S., \& Yee, J. 2014 spitz.prop 11006

\bibitem[Gould et al.(2015a)]{prop2015a} Gould, A., Yee, J., \& Carey, S. 2015a spitz.prop 12013

\bibitem[Gould et al.(2015b)]{prop2015b} Gould, A., Yee, J., \& Carey, S. 2015b spitz.prop 12015

\bibitem[Gould et al.(2016)]{prop2016} Gould, A., Carey, S., \& Yee 2016 spitz.prop 13005




\bibitem[Han \& Gould(1995)]{han95} Han, C. \& Gould, A.\ 1995, \apj, 447, 53

\bibitem[Henderson et al.(2016)]{henderson16} Henderson, C.B., Poleski, R., Penny, M. et al. 2016 \pasp 128, 124401



\bibitem[Kim et al.(2018)]{eventfinder} Kim, D.-J., Kim, H.-W., Hwang, K.-H., et al., 2018, \aj, in press

\bibitem[Kim et al.(2016)]{kmtnet} Kim, S.-L., Lee, C.-U., Park, B.-G., et al.  2016, JKAS, 49, 37







\bibitem[Refsdal(1966)]{refsdal66} Refsdal, S. 1966, \mnras, 134, 315


\bibitem[Ryu et al.(2017)]{ob161190} Ryu, Y.H, Yee, J.C., Udalski, A.\ et al. 2017, \aj, in press


\bibitem[Schechter et al.(1993)]{dophot} Schechter, P.L., Mateo, M., \& Saha, A. 1993, \pasp, 105, 1342







\bibitem[Shvartzvald et al.(2017)]{ukirt} Shvartzvald, Y., Bryden, G., Gould, A.\ et al.\ 2017, \aj, 153,61









\bibitem[Szyma\'nski et al.(2011)]{oiiicat}Szyma\'nski, M.K., Udalski, A., Soszy\'nski, I., et al. 2011, Acta Astron., 61, 83


\bibitem[Udalski et al.(1994)]{ews1} Udalski, A.,Szymanski, M., Kaluzny, J., Kubiak, M., Mateo, M.,  Krzeminski, W., \& Paczy\'nski, B. 1994, Acta Astron., 44, 227


\bibitem[Udalski et al.(2015)]{ob140124} Udalski, A., Yee, J.C., Gould, A., et al. 2015, \apj, 799, 237

\bibitem[Wo\'zniak(2000)]{wozniak2000} Wo\'zniak, P.~R. 2000, Acta Astron., 50, 421







\bibitem[Zang et al.(2018)]{cfht} Zang, W., Penny, M., Zhu, W.\ 2018, in prep





\bibitem[Zhu et al.(2017a)]{zhu17a} Zhu, W., Udalski, A., Calchi Novati, S.,  et al. 2017a, \apj, in press, arXiv:1701.05191

\bibitem[Zhu et al.(2017b)]{zhu17b} Zhu, W., Huang, C., Udalski, A., et al. 2017b, \pasp, 129, 104501 

\bibitem[Zhu et al.(2017c)]{zhu17c} Zhu, W., Udalski, A., Huang, C., et al. 2017b, \apjl, in press, arXiv:1709.09959

\end{thebibliography}
\end{document}